# Archaeoastronomy and the alleged "Stonehenge calendar"

Giulio Magli[1] and Juan Antonio Belmonte[2]

(1) Department of Mathematics, Politecnico di Milano, Italy.
(2) Instituto de Astrofísica de Canarias and Universidad de La Laguna, Tenerife, Spain.

*In a recent paper in Antiquity (Darvill 2022), the author has proposed that the project of the "sarsen" phase (stage 2) of Stonehenge (c. 2600 BC) was conceived in order to represent a calendar year of 365.25 days, that is, a calendar identical in duration to the Julian calendar. The aim of the present paper is to show that this idea is totally unsubstantiated, being based as it is on a series of forced interpretations and unsupported analogies.*

## 1. Introduction

Stonehenge is an astonishingly complex monument, which can be understood only by taking into account its landscape and the chronology of its different phases along the centuries (Gaffney et al. 2018, Pearson et al. 2021, 2022). There is, however, no doubt that the most spectacular and known phase, the "sarsen" one, exhibits a clear astronomical alignment which, due to the flatness of the horizon, refers both to the summer solstice sunrise and to the winter solstice sunset. This accounts for a clear, symbolic interest of the builders to the solar cycle, most probably related to the connections between afterlife and winter solstice in Neolithic societies (Ruggles 1997, 2014). However, this is, of course, very far from saying that the monument was used as a giant calendrical device, as instead has been proposed in a recent paper on Antiquity (Darvill 2022). In that paper, Stonehenge is interpreted as the representation of a calendar year of 365.25 days, that is, a calendar identical in duration to the Julian calendar, which properly entered in use only about two millennia later (González-García and Belmonte, 2007). The aim of the present letter is to show that this idea is unsubstantiated, being based as it is on a series of forced interpretations, numerology, and unsupported analogies with other cultures.

## 2. The alleged Stonehenge Calendar

In a nutshell, the "Stonehenge calendar" idea as presented in Darvill (2022) can be summarized as follows. The sarsen phase of the monument represents a calendar based on 365 days per year divided in 12 months of 30 days plus five epagomenal days, with the addition of a leap year every four. The number of days is obtained by multiplying the 30 sarsen lintels (probably) present in the original project by 12 (no 12 is recognizable in the monument) and adding to 360 the number of the standing trilithons of the Horseshoe, which is five. The addition of a leap year every four is related to the number of the station stones, which is, indeed, four. According to the author, the calendar was kept in operation using the solstitial alignment of the axis and was supposedly taken from Egypt. In doing this, however, the builders also undertook a refining from the Egyptian civil calendar because the leap year correction was not present there until Roman times.

We thus see that the proposal relies on 3 points:

1. A numerological argument
2. An Archaeoastronomy argument
3. An argument by analogy in cultural astronomy

Let us analyze the three points separately.

(1) **Numerology** is the pseudo-scientific way of reasoning that tends to attribute a meaningful, but hidden, relationship between numbers and concepts. These numbers are generated in various ways, from attributing numerical values to the letters of a text to extracting the measures of a building. This way of reasoning spans a range of significance, from acceptable to nonsensical (Dudley 1997). For instance, in the so-called Temple of the Inscriptions in Palenque, Chiapas, the foundations contain the tomb of the Mayan king Pakal, and the temple is nine-stepped as are the levels of the Maya underworld. This is a numerological observation which has a cultural basis and might be considered meaningful (Aveni 2015). At the opposite end, we can find in many fringe publications that the Angkor Wat temple in Cambodia sits 72° east of the Great Pyramid of Giza, "because" 72 are the years required to complete a degree of the precessional movement of the Earth. An assertion which, we believe, requires no comments at all.

Let us establish where exactly in the range of reasonableness sits the numerology of the recently proposed "Stonehenge calendar", as suggested by the author (Darvill 2022: 8). As mentioned, the proposal puts together the reconstructed number of the lintels of the external circle (30), the number of trilithons of the horseshoe (5), and the number of station stones (4). These 3 numbers lead the author to interpreting Stonehenge as a "calendar in stone", a sort of symbolic representation of a solar calendar identical to the Julian year as for the average length of the year (of course, as incorrectly stated, this length does not coincide with a tropical year which is slightly shorter). This calendar should be allegedly composed by 12 months (a number absent in the numerology exercise) of 30 days plus 5 epagomenal days (wrongly addressed as "intercalary month") plus one day every 4 years.

A first elaboration of this calendar is documented only two millennia later, when an attempt to put it in use was undertook in Ptolemaic Egypt (the so-called Canopus Decree) but apparently failed (Belmonte 2003, Hannah 2008). No such a calendar, thus, ever existed in due use before the reform of Julius Caesar of the Roman calendar (46 BC, González-García and Belmonte, 2007), and even this Julian calendar kept anomalous lengths for the months inherited from the Republican calendar tradition. Only later, under Augustus's reform, the Alexandrian calendar was developed in Egypt (c. 25 BC) with a similar structure (with an additional epagomenal day every four years) as the one proposed by the author but for the third millennium BC and in Britain.

Operationally – at least if we understand well the reasoning – at Stonehenge some device was used to sign the day of the 'month' represented by the corresponding stone in the trilithon circle and some unknown device was used on an unidentified set of 12 stones (or any other potential speculative tool) to sign the month. The device was then transferred to the standing trilithons in the epagomenal days; further, some device was used on the station stones to progressively count four years. Finally, no device (and/or stone) is proposed to add the extra day each four years.

As mentioned above, any tentative interpretation of "numbers in a monument" should be firmly grounded on a cultural basis, and this is even more true when astronomical observations come into play. However, before exploring this problem (point 3), it is clear from the very beginning that this "interpretation" suffers from two classical problems of Numerology: arbitrariness and the selection effect (e.g. Fagan 2006, and contributions therein). Arbitrariness is intrinsic in dividing the alleged calendar in 12 equal months of 30 days, since the number 12 is fully missing (the possibility that *"the poorly known stone settings in and around the north-eastern entrance somehow marked this cycle"* is awaken without any sort of evidence).

Selection is fully at play in discarding other numbers: the trilithon are made of 3 stones (sic), the portal was probably made of 2 (only the Heel stone remains). Besides this, there is an indeterminate number of "blue stones" that apparently did not play any role. Finally, the alleged month should have been divided in "decans" of 10 days (the correct word is of course "weeks" or "decades", because decans apply to a series of stars or asterisms in ancient Egypt used to keep track of nightly hours, see e.g. Neugebauer and Parker 1960, Belmonte 2002). The alleged proof for this decanal structure would be the difference in size of stones S11 and S21. However, something is going wrong here, as it is physically impossible that the small stone S11, as it is today (a break is speculated), sustained a lintel together with the much taller stone S10. Besides, almost half of the pillar stones of the trilithon circle have been lost and it is possible that they could have been small as well, thus breaking the magic of the hypothesis.

(2) The **Archaeoastronomy** (Magli 2020, Belmonte, 2021) argument is the following. The correctness and alignment of the alleged calendar with the solar cycle (anyway approximate, as mentioned the tropical year is slightly shorter than the average Julian year) was controlled by using the solstice alignment of Stonehenge. Some confusion is made here about the behavior of the sun at the solstice (the astronomical solstice is just the instant of maximal declination, and certainly not *"five days of standstill")* but the slow movement of the sun at the horizon makes it impossible to control the correct working of the calendar, as instead erroneously stated in the paper. This becomes more complicated the further north you go, especially in absence of a uneven reference horizon, because the sun rises less and becomes less perpendicular to the horizon itself. The consequence to this is that the Stonehenge alignment, although being accurate in space, cannot be accurate in time. In order to understand this point, let us suppose that we somehow know the correct day of the solstice in a certain year (a supposition which is, by itself, not easy to envision) and then we start counting periods of 365 days. In the fourth period, to tune the calendar we should be able to see that the solstice is occurring a day before, but in order to do this our device should be able to distinguish positions as accurate as a few arc minutes, a possibility which is clearly excluded. Therefore, we would employ many years to clearly notice the difference. Moreover, we would hardly be able to know how many days our calendar is going out of sync. As rightly stressed by one of our anonymous reviewers, this can be explained in less technical terms by stressing that the sarsen-phase solstitial axis is not accurate in time because of the minuscule difference between the sunrise/sunset position for several days by either side of either solstice. So, while the existence of the axis can be taken to demonstrate a 'calendrical' function in the very broad sense, the mere existence of the solstitial axis provides no proof whatsoever for inferring that the builders counted the days in the year and conceived it as comprising a set number of days, be that 360, 365 or 365.25.

(3) A third set of arguments relies on cultural astronomy **analogies**. This kind of arguments are delicate on their own, as trying to attribute to different cultures (especially distant ones) the same cognitive approach to astronomy is not exempt of risk. However, what happens here is even worse since the analogy in itself is flawed. The author says that this calendar makes a parallel with the Egyptian calendar (for an up-to-date approach to the Egyptian Archaeoastronomy and calendar the reader is referred to Belmonte 2012 and Magli 2013, and references therein).

There are several naive statements. For example, the starting date of the civil calendar is much debated, hence, it would be far-fetched to fix such a precise date as 2733 BC to its start . No Egyptologist would assume such an assertion as realistic (see Belmonte 2003, 2012, 27-48, for a discussion). Furthermore, the argument that the 12 months of the Egyptian calendar were named after the constellations that form the signs of the Zodiac demonstrates a poor knowledge on Egyptian astronomy (not only from the author but also from the side of the eventual reviewers and referees), indeed a crucial point. The relationship between the rise of the solar cult and the development of the civil calendar has been previously suggested but it is far from being demonstrated (Quirke 2001, Krauss 2011).

As we have argued, the Egyptian civil calendar was *not* the Julian calendar, but was a calendar made of strictly 365 days. Therefore, it rapidly went out of sync – it was *not* a solar calendar after all, this is why it is often called a "wandering calendar". As the author recalls, they probably knew this or detected this problem after years of using their "wandering calendar", but they, very simply, did not feel the necessity of changing the calendar itself up to "much later" (which means some 27 centuries).

They perhaps reflected the drift of the calendar through the seasons in their architecture (see e.g., Belmonte 2009), presumably in a symbolic manner, but we have no evidence that they ever erected a monument to control time. Hence, the argument that materializing *a time-reckoning system in the structure and form of a mayor monument ... is common practice amongst non-literate and semi-literate societies ...* is surprising. In 30 years of work in the field, we have never found an unmistakable proof of this affirmation before the arrival of scientific astronomy and its true observatories. Even more so, this has not been found in Egypt, the supposed source of inspiration , unless we are willing to accept an independent development in the third millennium BC Britain. Indeed, we have not a single piece of evidence to support that (Belmonte 2015, Ruggles and Cotte 2017).

In the mind of the author, the Stonehenge people not only received information of a 365 days calendar, but also knowledge by the Egyptian astronomers that their calendar was not anchored with the solar cycle, and decided to "cure" this in a way similar (if not better) to the Roman astronomers in Julius Caesar's times. A transfer and elaboration of notions with Egypt (occurred around 2600 BC) which indeed resembles very old ideas of culture diffusionism and the like – in particular, the solar worshipers' travels in antiquity is an old Thor Heyerdahl's paradigm that we believed had been discarded and happily forgotten. Besides, as mentioned in the paper, there is only a one-off case of a very far visitor to Stonehenge – who was, anyhow, arriving from the Alps – and one single debated case of one object's provenience (dated from a millennium later than Stonehenge S2). At least in our view, this is far from showing that "*the pendulum of interpretation is swinging back in favor of long-distance contacts and extensive social networks*" (Darvill, 2022). Using the alleged calendar as a proof, as done in the paper under exam, is, of course, a circular argument. Equally debatable is awaking to the vague references present in medieval accounts written some 3600 years after construction.

## 3. Conclusions

All in all, we have shown here that the alleged "Neolithic-Julian" Stonehenge calendar is a purely modern construct whose archaeoastronomical and calendrical bases are flawed. Interestingly, from a strictly cultural point of view, the justification for a solar-anchored calendar at Stonehenge is also not as well sustained as the author apparently imagines. Most evidence suggests that early societies, including Neolithic ones, presumably used lunisolar calendars (Belmonte et al. 2019, Belmonte 2021), as it has been done by most people on Earth in antiquity and even today (Stern 2012), with the exceptions of the ancient Egyptians (and Romans by influence) and the Maya in America. Having more or less precise solar alignments could perhaps be used to anchor lunar New Year's Eves, but would hardly be sufficient to develop an operative solar calendar. For this, before the telescope, one would have needed devices as precise as the sundial present in the Jantar Matar in Jaipur. Stonehenge is, evidently, not such a device!

We understand that publishing "high-risk/high-performance" results is tempting, but Archaeoastronomy had to endure decades' long of difficult development in order to become the respected scientific discipline that it is today (see e.g. Aveni 2008, Magli 2020, Boutsikas et al. 2021). We believe that matters such as ancient calendars, alignments and cultural astronomy should be reserved to specialists, to experienced people who have trained in the adequate dominion on the subject, and not left to enthusiasts, even if those same enthusiasts are renowned and knowledgeable specialists in their own field. Perhaps, multidisciplinarity and a bit of common sense are still the best scope to have in mind.

**Acknowledgements.** The excellent insight of two anonymous reviewers has greatly improved the quality of the manuscript. JAB acknowledges support from the State Research Agency (AEI), Spanish Ministry of Science, Innovation (MICIN) and the European Regional Development Fund (ERDF) under grant with reference AYA2015-66787-P "Orientatio ad Sidera IV" and under the internal IAC project P310793 "Arqueoastronomía."